\documentstyle[prl,aps,epsfig]{revtex}         % 2 col mode
\begin{document}
\draft               % preprint mode
% 2 col mode:
\twocolumn[\hsize\textwidth\columnwidth\hsize\csname @twocolumnfalse\endcsname

\title{From coherent to incoherent hexagonal patterns in semiconductor resonators }
\author{V. B. Taranenko, C. O. Weiss, B. Sch\"{a}pers$^*$}
\address{Physikalisch-Technische Bundesanstalt 38116 Braunschweig, Germany\\
$^*$Westf\"{a}lische Wilhelms-Universit\"{a}t, 48149 M\"{u}nster,
Germany}
%\date{\today{}}
\maketitle
\begin{abstract}
We find that hexagonal structures forming in semiconductor
resonators can range from coherent patterns to arrangements of
loosely bound spatial solitons, which can be individually
switched. Such incoherent arrangements are stabilized by gradient
forces, as evidenced by the stability of hexagonal structures
with single- or multiple-soliton defects. We interpret the
experimental observations by numerical simulations based on a
model for a large aperture semiconductor microresonator.
\end{abstract}
 \vskip1pc ]
% BEGIN TEXT HERE
Pattern formation in the form of hexagonal structures was
predicted years ago for resonators containing a reactive
nonlinearity \cite{tag:1}. We have recently given the first proof
of the phenomenon using semiconductor microresonators in the
dispersive limit \cite{tag:2}. This kind of pattern formation was
regarded as an important precursor of optical soliton formation,
the latter being of technical importance for all-optical
information processing \cite{tag:3}. We showed the existence of
these bright and dark spatial solitons recently
\cite{tag:4,tag:5}.

Our finding, that individual bright spots of the hexagonal
patterns can for certain parameters be "switched off" without
apparent effects for the rest of the hexagonal pattern
\cite{tag:2}, caused a debate about whether such individually
switchable "pixels" in a hexagonal pattern have soliton
properties and more in general about the nature of these
hexagonal patterns.  In this article we clarify these questions
by interpreting our experimental observations using numerics on a
semiconductor resonator model \cite{tag:6}.\\

For completeness we repeat shortly the experimental arrangement:
Light of wavelength near the semiconductor band edge (850 nm),
generated by a continuous Ti:Al$_2$0$_3$-laser, irradiates an
area of 50-100 $\mu$m diameter of the semiconductor resonator
sample, with intensity of up to 3 kW/cm$^{2}$ . The sample is a
quantum-well stack between Bragg mirrors of 99.7 $\%$ reflectivity
\cite{tag:7}. The optical resonator length is about 3 $\mu$m so
that a Fresnel number of several 100 is excited, sufficient for
complex structure to form. The light is admitted to the sample
for durations of a few $\mu$s (through a mechanical chopper, to
limit thermal phenomena) repeated every ms.

As the substrate, on which the resonator sample is grown, is
opaque at the wavelength used, all observations are done in
reflection. Either by taking ns-snapshots of the illuminated
area, or by following the reflected intensity in particular
points of the illuminated area as a function of time, using a
fast photodiode. Details are given in \cite{tag:2,tag:5}.\\

Fig.~1 shows structures observed at a wavelength of 880 nm (i.e.
to a good approximation in the defocusing Kerr limit), for
different illumination intensities and detunings
$\delta$$\lambda$ of the light from the resonator resonance
center. The half-width-half-maximum of the resonance is $\approx$
0.1 nm. At a detuning of -0.1 nm, for lower intensities a
hexagonal pattern of dark spots forms (Fig.~1a). This converts to
a hexagonal pattern of bright spots for higher illumination
(Fig.~1b).

\begin{figure}[htbf] \epsfxsize=60mm
\centerline{\epsfbox{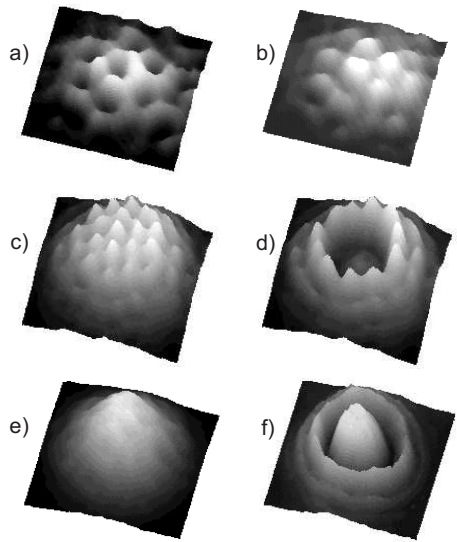}} \vspace{0.7cm} \caption{Intensity
structures (in 3D representation) observed in reflection for
different resonator detuning $\approx$ -0.1 nm (a, b), -0.2 nm
(c, d) and -0.4 nm (e, f). Intensity of incident Gaussian beam is
increased from top to bottom and from left to right.}
\end{figure}

For $\delta$$\lambda$ $\approx$ -0.2 nm at lower intensity a
bright hexagon forms (Fig.~1c). The smaller period at
$\delta$$\lambda$ $\approx$ -0.2 nm compared to $\delta$$\lambda$
$\approx$ -0.1 nm indicates that these patterns are formed by the
"tilted-wave"-mechanism \cite{tag:8} as in most known cases of
pattern formation in optical resonators \cite{tag:1,tag:9}. At
$\delta$$\lambda$ $\approx$ -0.2 nm the resonator characteristic
is already bistable so that the central part of the illuminated
area is switched at a larger intensity as shown in Fig.~1d. The
switched area is the central part of the Gaussian illumination
beam in which the intensity exceeds the "Maxwellian intensity"
\cite{tag:10}, the latter being the intensity at which the
switching front surrounding the switched area does move neither
radially inward, nor outward.

At $\delta$$\lambda$ $\approx$ -0.4 nm the structure is no longer
visible (Fig.~1e). The pattern period at this large detuning is
small enough to apparently be washed out by non-local effects
like carrier diffusion. The switching of the central part at
higher intensity then has the appearance of switching in an
unstructured environment (the counterintuitive central intensity
peak in case (f) compared with case (d) results due to the
nonlinear resonance effect). We note the radial modulation
outside the switched area, clearly apparent for the switched
cases Figs 1d and 1f (the azimuthal modulation of the first ring
in Fig.~1d is a residual of the hexagonal structure in Fig.~1c).
This radial modulation is due to the "oscillating tails" of the
switching front \cite{tag:10} responsible for such phenomena as
stabilization of solitons \cite{tag:10}, forming bound states of
several solitons ("molecules") \cite{tag:11}, or stabilizing
large patterns (as shown below), through the forces associated
with the gradients of the modulations \cite{tag:12}.

\begin{figure}[htbf]
\epsfxsize=85mm \centerline{\epsfbox{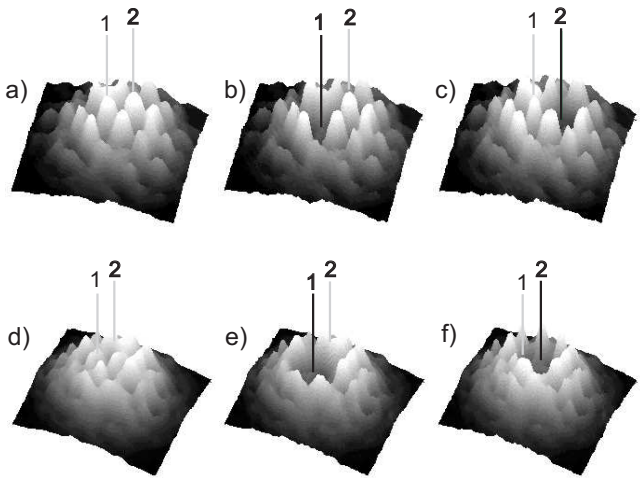}} \vspace{0.7cm}
\caption{Switching of individual spots of hexagonal structures
with focused address pulses aimed at different places (marked 1
and 2) of the pattern. See text for details.}
\end{figure}

We note for clarity that the switching in the lower two rows of
Fig.~1 is effected by increasing the intensity of the (Gaussian)
illumination beam. As opposed to that in Fig.~2 switching is done
by an additional more sharply focused beam (of perpendicular
polarization, thus incoherent with the main illumination field).
For the switching experiments the additional light, focused to a
diameter of 8 $\mu$m, comparable in size to the bright spots in
the hexagonal patterns of Fig.~1 and Fig.~2, was aimed at
particular ones of the bright spots in the hexagonal patterns in
short pulses (10 ns).

Fig.~2 a, b, c shows two bright spots (marked 1 and 2) at which
the switching light pulses were aimed. In this way pixel 1 is
switched off in Fig.~2b while all other pixels of the hexagonal
pattern remain unaffected. Likewise, aiming the switching pulse
at pixel 2 switches this latter one off without affecting the
rest of the pattern (Fig.~2c).

We recall that we speak here of switching in the strict sense:
pixels 1 and 2 in Fig.~2b, ~2c, respectively remain switched off,
stationarily, after the end of the short (10 ns) switching pulse.
It was also possible, as Figs~2 d, e, f show, to switch several
pixels at once by using a higher intensity of the switching beam
and aiming at a spot surrounded by 3 pixels. Figs~2 d, e, f show
that different "pixel-triples" can be switched when the switching
beam is aimed at different locations.

The "local switching" results of Fig.~2 suggest that the
hexagonal patterns we observe are not necessarily "coherent
patterns" in the sense that a perturbation in one part of a
pattern affects the entire pattern. Rather in Fig.~2 the
hexagonal patterns appear to behave like a collection of loosely
bound individual spatial solitons whose structure and stability
is independent of the rest of the pattern. (And vise versa the
rest of the pattern is unchanged if individual solitons are
removed). On the other hand, for other parameters such as for
Fig.~1 c, d, there are coherent patterns in which a local
perturbation with focused address pulses aimed at different
places leads to a destruction of the whole pattern.

In order to gain more insight into the stability/instability of
the various structures we investigated numerically the model
\cite{tag:6} for a large aperture semiconductor resonator, which
we found in good agreement with experimental results recently
\cite{tag:13,tag:14}. The set of equations for the intracavity
field $E$ and the carrier density $N$ is:
\begin{equation}
\cases{{\partial E}/{\partial t}=E_{\rm in}-E[1+C{\rm
Im}(\alpha)(1-N)]-\cr
\quad\quad\quad\quad\quad\quad\quad\quad-iE[\theta-C{\rm
Re}(\alpha)N-\nabla^{2}_{\bot}]\,,\cr \cr{\partial N}/{\partial
t}=-\gamma[N-|E|^2(1-N)-d\nabla^{2}_{\bot}N]\,,}
\end{equation}
where $E_{\rm in}$ is the incident field, $C$ is the bistability
parameter \cite{tag:6}, ${\rm Im}(\alpha)(1-N)$ and ${\rm
Re}(\alpha)N$ describe the absorptive and refractive
nonlinearities, respectively. $\theta$ is the detuning of the
optical field from the resonator resonance, $\gamma$ is the ratio
of the photon lifetime in the resonator to the nonradiative
carrier recombination time, $d$ is the diffusion coefficient
(normalized to the diffraction coefficient) and
$\nabla^{2}_{\bot}={\partial^{2}\nonumber}/{\partial
x^{2}}+{\partial^{2}\nonumber}/{\partial y^{2}}$ is the
transverse Laplacian.

Fig.~3 shows the calculated resonator plane wave characteristic
for a rather dispersive nonlinearity (Kerr-type, defocusing). We
note that this characteristic is not usually stable (i.e. plane
waves are not necessarily stable solutions). For a large detuning
the lower branch is generally modulationally unstable (dashed
line Fig.~3). At smaller intensity (Fig.~3a) a bright spot
hexagon develops supercritically, whereas it becomes a dark spot
hexagon at higher intensity (Fig.~3c,d). Transition to patterns
at higher intensity is subcritical because existence ranges for
stable structures overlap. The bright hexagons transform to the
dark hexagons via a stripe pattern (Fig.~3b) formed at
intermediate intensity.

\begin{figure}[htbf]
\epsfxsize=80mm \centerline{\epsfbox{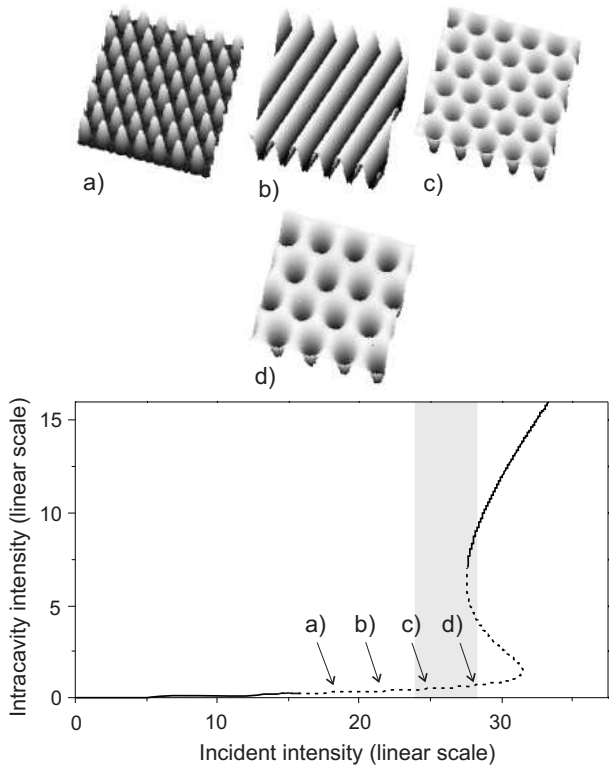}} \vspace{0.7cm}
\caption{Numerical solutions of Eq.~(1) for intracavity intensity
versus incident intensity:  homogeneous solution (dashed line
marks unstable part of the curve) and patterns (a - d). Shaded
area marks existence range for dark spot hexagons. Parameters: $C
= 100$, $\rm Im(\alpha)=0.01$, $\rm Re(\alpha)=0.1$,
$\theta=-10.3$, $\gamma =
0.1$, $d = 0.01$.\\
Dark/bright spots correspond to bright/dark spots in the
experiment where observation is in reflection.}
\end{figure}

We note the different periods of the structures in Fig.~3. As can
be seen, the pattern period is increased from a) to d) which
translates into smaller wavefront tilt, indicating a smaller
effective detuning, whereas the external resonator detuning is
the same for all cases. The mechanism of importance here is
"nonlinear resonance" \cite{tag:15}, i.e. the internal parameter
of field and with it the nonlinear refraction adjust to reduce
the detuning (and the wavefront tilt). Consequently, with the
higher input intensity, for example in case d), the structure
formation is of more nonlinear origin than  in case c). In turn,
case c) is of more nonlinear origin than cases b) and a). Thus
the cases a) and c) represent patterns governed by the tilt of 6
plane phase-locked waves, the tilt being prescribed by the
external parameter of resonator detuning. The case d) is not
exclusively determined by this external parameter but the high
nonlinearity gives the system more internal freedom and
flexibility permitting larger numbers of stable patterns.

Numerical experiments show that this is indeed the case. Fig.~4
shows that it is possible to remove, as in the experiment, one
dark spot, or three dark spots, from the patterns without
destabilizing the rest of the structure. One sees in Fig.~4 b, c
that at the locations of the removed solitons, the field is not
uniform. The residual field nonuniformities are the result of the
superposition of the "oscillating tails" surrounding all dark
spots. For the case Fig.~3d, one could therefore picture the
structure as a weakly bound collection of dark solitons. In the
hexagonal arrangement, the "oscillating tails" of all solitons
surrounding one particular site superpose at this site to a field
nonuniformity which can trap a soliton. And, vice versa, each
soliton contributes a field at the sites of all its neighbours
which stabilizes the latter's positions.

Structures with patches of more than 3 solitons missing are also
found stable. In general the larger the number of solitons
missing, the smaller the range of stability of the pattern. It
can thus be said that the lower branch of Fig.~3 in the vicinity
of d) is not modulationally unstable but unstable against
formation of collections of dark spatial solitons with hexagonal
geometry and defects. This includes fields with few bound or
isolated spatial solitons (Fig.~5). Many different forms of such
soliton arrangements coexist. The hexagonal matrices as in
Fig.~2a can therefore carry substantial amounts of information -
as may be useful in applications for parallel optical information
processing.

\begin{figure}[htbf]
\epsfxsize=80mm \centerline{\epsfbox{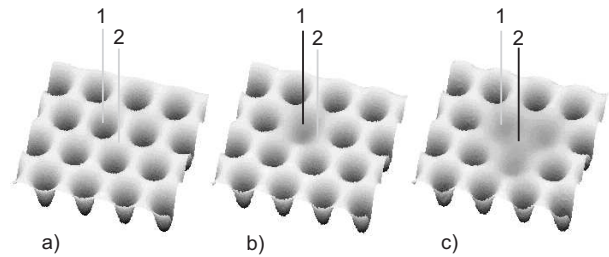}} \vspace{0.7cm}
\caption{Stable hexagonal arrangements of dark spatial solitons:
a) without defects, b) with single-soliton defect and c) with
triple-soliton defect. Numerical integration of Eq.~(1) for
$E_{\rm in}=5.30$ and other parameters as in Fig.~3.
Locations "1" and "2" are where the switching pulses are aimed.\\
For correspondence with experiment see remark in caption Fig.~3.}
\end{figure}

In this numerical interpretation of the experimental results we
have to add, however, a moment of caution: Although the numerics
reproduces the observations quite well and completely, the
detuning parameters for which we find in the model the phenomena
observed experimentally, are much larger than in the experiment.
Our suspicion that the switched pixels might result from another
mechanism than solitonic localization, however, was not supported
by further model calculations. We tested the hypothesis that the
"localized switching" as observed might result in the following
alternative way: A hexagonal modulation (as in Fig.~1 and 2)
develops spontaneously on the lower branch starting from
scattered light spatially filtered by the detuned resonator
\cite{tag:2}. When switching locally, the switched area is
limited in size, stabilized and held in one place by gradients of
the spatial modulation. An extensive numerical search for such
behavior (at small detuning corresponding to the experimental
parameters) failed. To our understanding this should rule out an
alternative mechanism differing from solitonic localization. The
discrepancy between the detunings in experiment and model
calculation is probably explained such that due to material
effects, as heating, the detuning during the observations was
effectively larger than that taken in absence of radiation.

The picture of formation of hexagonal patterns in nonlinear, and
particularly dispersive, resonators is in general perceived as
that of emission of 6 tilted phase-locked waves pumped by a
4-wave-mixing process between the illumination field and the
generated fields. Thus it resembles laser emission, where the
tilt of the generated waves is forced by the resonator detuning
\cite{tag:8}.

\begin{figure}[htbf] \epsfxsize=80mm
\centerline{\epsfbox{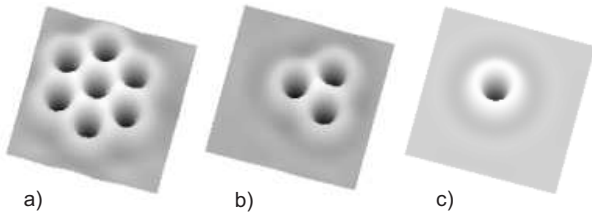}} \vspace{1.0cm} \caption{Dark
solitons on homogeneous background obtained by sequentially
removing individual elements from an incoherent hexagonal
structure like in Fig.~4 : a) seven and b) three bound dark
solitons, c) single dark soliton. Numerical integration of
Eq.~(1) for $E_{\rm in}=5.27$ and other parameters as in
Fig.~3.\\
We note that for intensity corresponding to Fig.~4 ($E_{\rm
in}=5.30$) single solitons and bound soliton pairs are unstable;
and molecules of more than 2 solitons are all stable.}
\end{figure}

This picture describes formation of coherent hexagons. We find
here, however, that the hexagonal structures observed have the
properties of "densest packed" individual spatial solitons, which
are loosely bound. This implies that there should exist a
continuous transition from extended coherent patterns to
collections of independent spatial solitons. The parameter
governing this transition is increasing nonlinearity. To
illustrate, Fig.~3a,b,c would be coherent patterns, while with
increasing input intensity the loosely bound densest packed
solitons Fig.~3d develop.\\

Concluding, we find here that hexagonal structure formation as we
observe it in semiconductor microresonators can lead to
coherent-extended patterns, as well as to "crystals" of bound
spatial solitons. Such matrices of solitons can therefore carry
substantial amounts of information in applications for optical
parallel processing.

Similar effects have been observed in other nonlinear optical
systems with feedback \cite{tag:11,tag:16}. We would therefore
conclude that the mechanism which changes the nature of the
hexagonal patterns from coherent to incoherent is a rather general one.\\

Acknowledgement\\ This work was supported by the ESPRIT LTR
project PIANOS. We gratefully acknowledge discussions with W.
Lange and T. Ackemann in the frame of the ESF network PHASE. We
thank K. Staliunas for help with the numerics.

\end{document}